\documentclass[aps,prl,twocolumn,superscriptaddress,floatfix]{revtex4}
\usepackage{graphicx}
\usepackage[]{amsmath}

\begin{document}

% number in the upper right hand corner of the title page in preprint mode.
%\preprint{First Draft}

\title{The phase transition in the localized ferromagnet EuO probed
by $\mu$SR}


\author{S. J. Blundell}
\email{s.blundell@physics.ox.ac.uk}
%\affiliation{Clarendon Laboratory, University of Oxford, Parks Road, Oxford OX1 3PU, United Kingdom}
\author{T. Lancaster}
\affiliation{Clarendon Laboratory, University of Oxford, Parks Road, Oxford OX1 3PU, United Kingdom}
\author{F. L. Pratt}
\affiliation{ISIS Muon Facility, ISIS, Chilton, Oxon. OX11 OQX, United Kingdom}
\author{P. J. Baker}
\affiliation{Clarendon Laboratory, University of Oxford, Parks Road, Oxford OX1 3PU, United Kingdom}
\affiliation{ISIS Muon Facility, ISIS, Chilton, Oxon. OX11 OQX, United Kingdom}
\author{W. Hayes}
\affiliation{Clarendon Laboratory, University of Oxford, Parks Road, Oxford OX1 3PU, United Kingdom}
\author{J.-P. Ansermet}
\author{A. Comment}
\affiliation{Institut de Physique
de la Mati\`ere Condens\'ee, 
Ecole
  Polytechnique, F\'ed\'erale de Lausanne, CH-1015 Lausanne-EPFL, Switzerland}

%\thanks{}
%\affiliation{}


\date{\today}

\newcommand{\chem}[1]{\ensuremath{\mathrm{#1}}}

\begin{abstract}
We report results of muon spin rotation measurements performed on the
ferromagnetic semiconductor
EuO, which is one of the best approximations to a localized
ferromagnet.
We argue that implanted muons are sensitive to the internal field primarily
through a combination of hyperfine and Lorentz fields.  The
temperature dependences of the internal field and the relaxation rate
have been measured and are compared with previous theoretical predictions.
\end{abstract}

%\keywords{}

\maketitle


Europium oxide (EuO) crystallises in the rock-salt structure and is a
ferromagnetic semiconductor with a Curie temperature ($T_{\rm C}$) of
69\,K \cite{matthias}.  It shows a colossal magnetoresistance effect
in its Eu-rich form \cite{shapira,nolting} and the associated
metal-insulator transition has been linked to the formation of bound
magnetic polarons \cite{torrance}.  It is the only magnetic binary
oxide known to be thermodynamically stable in contact with silicon
\cite{hubbard}, and this, together with a nearly 100\% spin
polarization of mobile electrons for carrier concentrations below half
filling of the conduction band \cite{sattler,steeneken}, means that it
is thought to be highly relevant for spintronic applications
\cite{nmat}.  The magnetic moments of the Eu$^{2+}$ (4f$^7$,
$^8$S$_{7/2}$) ions result from 4f charge density which is nearly
completely localized inside the filled 5s$^2$5p$^6$ shells and there
is negligible overlap. This makes EuO an excellent approximation to a
Heisenberg ferromagnet \cite{passell}, although there is evidence of
some momentum dependence in the exchange interactions \cite{miyazaki}.
The ferromagnetic interactions are due to an indirect exchange
mediated by electrons in anion valence bands \cite{leeliu},
specifically virtual excitations of oxygen valence band p electrons
into empty Eu$^{2+}$ (5d) conduction bands and exchange interaction of
the d electron (p hole) with the localized 4f electrons.  EuO is one
of a family of isostructural europium chalcogenides EuX with X=O, S,
Se and Te.  As the size of the chalcogen increases from O to Te, the
lattice parameter $a$ increases across the series: 5.14\,\AA\ in EuO,
5.97\,\AA\ in EuS, 6.20\,\AA\ in EuSe and 6.60\,\AA\ in EuTe
\cite{euoreview}.  Only EuO and EuS are ferromagnets and both the
nearest-neighbor and next-nearest-neighbour exchange constants are
larger in EuO compared to EuS \cite{passell,bohn}.  In EuO, the
saturation magnetization $M_{\rm s}$ is given by $gJ\mu_{\rm
  B}(4/a^3)$, yielding $\mu_0M_{\rm s}=2.40$\,T (using $gJ=7$).

Neutron studies of EuO are significantly hindered by the strong
absorption of neutrons by Eu, though the use of a thin-slab geometry
and use of $^{153}$Eu (which has the smaller absorption cross section
of the two naturally occurring isotopes) has allowed a detailed study to
be performed \cite{passell}.  Nevertheless, a technique such as
muon-spin rotation ($\mu$SR) avoids these difficulties entirely, and
in contrast to NMR there is no electric field gradient or quadrupolar
contribution to the observed muon response, simplifying the analysis.
In this paper we present the results of $\mu$SR experiments to study
the magnetic order and the fluctuations in samples of EuO and use the
results to compare with theoretical predictions
\cite{lovesey1,lovesey3,yaouanc}.

\begin{figure}
\includegraphics[width=8.5cm]{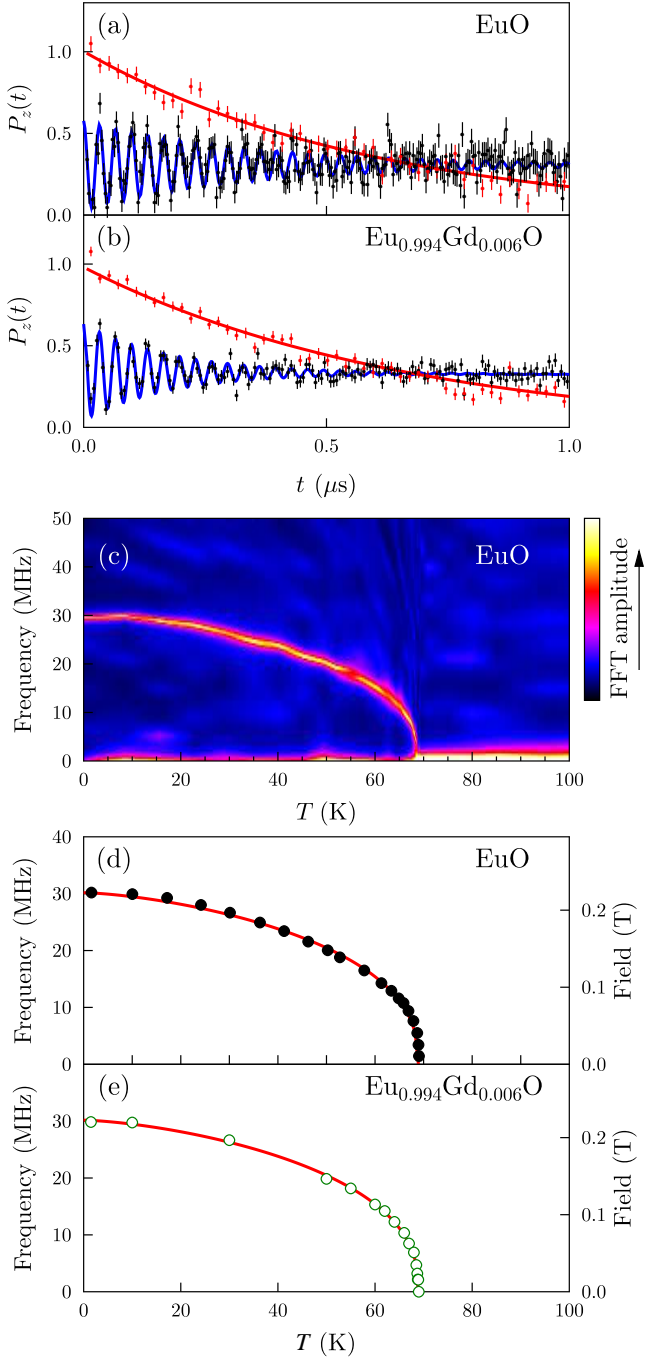}
\caption{(Color online.) (a) Raw $\mu$SR data for EuO.  Above $T_{\rm
    C}$ (at 70\,K), $P_z(t)$ relaxes. An oscillating signal develops
  below $T_{\rm C}$ (data shown for
  1.5\,K). (b) The same for  Eu$_{0.994}$Gd$_{0.006}$O.
  (c) Plot showing FFT of the muon data for EuO as a function of
  temperature.  The single precession frequency which follows the
  order parameter is clearly visible.
  (d) The extracted precession frequency for EuO 
   as a function of temperature.  
  (e) The same experiment repeated for Eu$_{0.994}$Gd$_{0.006}$O.
The line through the data in (d) and (e) is identical for the two
samples
and is a best fit of the data to the phenomelogical function 
$\nu(T)=\nu(0)(1-(T/T_{\rm C})^\alpha)^\beta$, producing
$\alpha\approx 1.5$, $\beta\approx 0.4$). A more reliable
extraction of the critical exponent $\beta$, focussing only on the
critical regime, is given later in the paper.
\label{fig:euo-raw}}
\end{figure}

Our $\mu$SR experiments were carried out using the GPS instrument at
the Swiss Muon Source (S$\mu$S), Paul Scherrer Institute (PSI) in
Switzerland.  In our $\mu$SR experiment, spin polarised positive muons
($\mu^+$, momentum 28~MeV$/c$) were implanted into small crystals of
EuO. The muons stop quickly (in $<10^{-9}$~s), without significant
loss of spin-polarization. The observed quantity is then the time
evolution of the average 
muon spin polarization $P_z(t)$, which can be detected by
counting emitted decay positrons forward (f) and backward (b) of the
initial muon spin direction; this is possible due to the asymmetric
nature of the muon decay \cite{musr,dalmas}, which takes place in a mean time
of $2.2~\mu{\rm s}$.  In our experiments positrons are detected by
using scintillation counters placed in front of and behind the sample.
We record the number of positrons detected by forward ($N_{\rm{f}}$)
and backward ($N_{\rm{b}}$) counters as a function of time and
calculate the asymmetry function, $G_{z}(t)$, using
$G_{z}(t)=[N_{\rm{f}}(t)-\alpha_{\rm exp}
N_{\rm{b}}(t)]/[N_{\rm{f}}(t)+\alpha_{\rm exp} N_{\rm{b}}(t)]$,
where $\alpha_{\rm exp}$ is an experimental calibration constant and
differs from unity due to non-uniform detector efficiency. The
quantity $G_{z}(t)$ is then proportional to $P_z(t)$.

Raw data for EuO at two temperatures, one slightly higher and one much lower
than $T_{\rm C}$, are shown in Fig.~\ref{fig:euo-raw}(a).  These
demonstrate that spin relaxation above $T_{\rm C}$ changes to a damped
coherent oscillation below $T_{\rm C}$.  If the experiment is repeated
with a sample of EuO in which 0.6\% of the Eu ions are replaced with
Gd, an almost identical result is obtained [Fig.~\ref{fig:euo-raw}(b)]
although the damping rate of the oscillations in the ordered state is
noticeably larger (by about a factor of two).  
This level of Gd doping is known to introduce a
static magnetic inhomogeneity \cite{comment}.  The frequency of the
oscillations in both cases increases rapidly on cooling below $T_{\rm
  C}$, see Fig.~\ref{fig:euo-raw}(c,d,e), approaching $\approx
30$\,MHz at zero temperature.  The Fast Fourier transform data in
Fig.~\ref{fig:euo-raw}(c) demonstrate that only a single precession
frequency is observed.  The temperature dependence of the precession
frequency in the Gd-doped sample is almost identical to that in the
pure sample, demonstrating that the order parameter is following the
intrinsic magnetism in the EuO host and is relatively insensitive to
low levels of doping.  In fact, at the 0.6\% level of Gd-doping, 93\%
of the Eu ions have their full complement (twelve) of
nearest-neighbour Eu ions, with most of the remaining Eu ions having
only one of those nearest neighbours replaced by Gd.
For both
samples, the relaxation rate of the oscillatory component rises as $T$
approaches $T_{\rm C}$ from below and the low-temperature
relaxation rate of the Gd-doped sample is larger than that of the pure sample

\begin{figure}
\includegraphics[width=8.0cm]{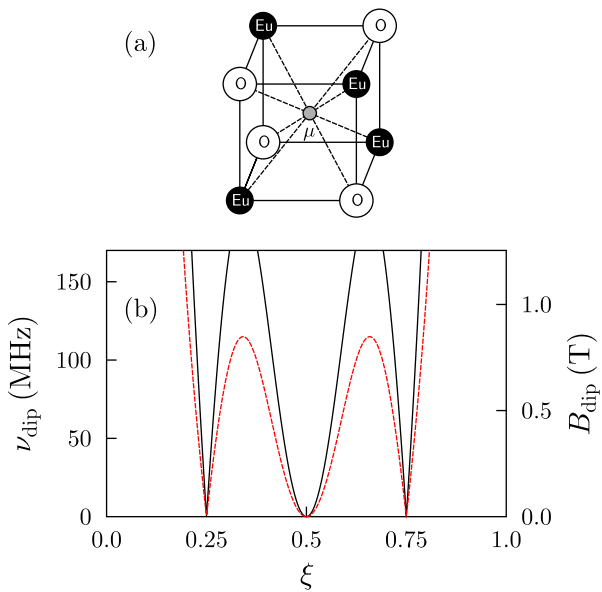}
\caption{(Color online.) (a) Muon site in EuO.  The crystal structure for one-eighth
  of the unit cell is shown (the side of the cube is $a/2$).  (b) Calculated dipole field along the
$\langle 111\rangle$ axes in EuO. The muon is located at $\xi\xi\xi$ and the Eu moments
are all aligned along [111] (solid black line) or one of [$\bar{1}$11],
[1$\bar{1}$1] or [11$\bar{1}$] (dashed red line).
\label{fig:fieldplot}}
\end{figure}

The muon spin precesses around a local
magnetic field, $B_\mu$ (with a frequency $\nu=(\gamma_{\mu}/2\pi) \vert B_\mu
\vert$, where $\gamma_{\mu}/2\pi= 135.5~\mathrm{MHz\,T}^{-1}$).  This
local field ($B_\mu \approx 0.22$\,T at $T=0$) 
is a sum of various terms, including the Lorentz field $B_{\rm L}$,
the hyperfine field $B_{\rm hf}$, the demagnetizing field $B_{\rm
  demag}$ and the dipolar field ${\bf B}_{\rm dip}$.
The latter quantity is a function of the muon-site
${\bf r}_\mu$ and can be written as
\begin{equation}
B_{\rm dip}^\alpha({\bf r}_\mu) = \sum_i D_i^{\alpha\beta}({\bf
r}_\mu)\,m_i^\beta, 
\label{eq:localfield}
\end{equation}
a sum over the magnetic ions in the crystal; the
magnetic moment of the $i$th ion is ${\bf m}_i$.  
In Eq.~(\ref{eq:localfield}), 
$D_i^{\alpha\beta}({\bf r}_\mu)$ is
the dipolar tensor given by
%\begin{equation}
$
D_i^{\alpha\beta}({\bf r}_\mu) = {\mu_0\over 4\pi R_i^3} \left(
{3 R_i^\alpha R_i^\beta \over R_i^2} - \delta^{\alpha\beta} \right)$,
%\end{equation}
where ${\bf R}_i\equiv (R_i^x,R_i^y,R_i^z)={\bf r}_\mu-{\bf r}_i$.
The behaviour of this tensor is dominated by the arrangement of the
nearest-neighbour magnetic ions and leads to a non-zero local magnetic
field for almost all possible muon sites \cite{musr08}.  On
electrostatic grounds, the likely muon site in EuO is at the
$\frac{1}{4}\frac{1}{4}\frac{1}{4}$ position [see
  Fig.~\ref{fig:fieldplot}(a)], equidistant from four Eu cations and
four oxygen anions.  Because of the magnetic anisotropy \cite{kasuya},
the easy axis for the Eu moments is along the $\langle 111\rangle$ set
of directions and for this moment alignment the
$\frac{1}{4}\frac{1}{4}\frac{1}{4}$ position is a point at which the
dipolar magnetic field actually vanishes.  The value of $B_{\rm dip}$
has been calculated for the case in which the muon is displaced from
the $\frac{1}{4}\frac{1}{4}\frac{1}{4}$ site and its position is
allowed to vary along the [111] direction, see
Fig.~\ref{fig:fieldplot}(b).  The dipolar field vanishes at both
$\frac{1}{4}\frac{1}{4}\frac{1}{4}$ (muon site) and
$\frac{1}{2}\frac{1}{2}\frac{1}{2}$ (oxygen site) and increases
sharply as the site moves away from these special positions of high
symmetry.  The two curves show the cases in which the muon
displacement is the same or a different choice of $\langle 111
\rangle$ direction.  In the former case, the dipolar field at the muon
site is parallel to the moment direction; in the latter case (for
which there are three possibilities), it lies along one of the
crystallographic axes and its amplitude is reduced by a factor of
$\sqrt{3}$.  Thus if the muon site is displaced from the
$\frac{1}{4}\frac{1}{4}\frac{1}{4}$ position towards a particular
oxygen anion, then there would be a contribution to the dipole field
resulting in two precession frequencies with a ratio of $\sqrt{3}$,
and with amplitudes in a 3:1 ratio.  Since only a single frequency is
observed, we conclude that the muon site is indeed at the
$\frac{1}{4}\frac{1}{4}\frac{1}{4}$ position in which the dipolar
field is zero, so that the observed local field ($B_\mu = 0.22$\,T at
$T=0$) is due to a sum of the Lorentz field ($B_{\rm L} = \mu_0
M/3=0.80$\,T at $T=0$), $B_{\rm demag}$ and $B_{\rm hf}$. Since the
sample is polycrystalline and multidomain, we neglect $B_{\rm demag}$
and deduce that the hyperfine field $B_{\rm hf}<0$ (antiparallel to
the magnetization), as found for EuS \cite{eschenko}, and takes the
value $B_{\rm hf} = -B_{\rm L}\pm B_{\rm \mu}$, and so either $-0.58$
or $-1.02$\,T.  For both samples, the amplitude of the oscillatory
component is reduced from the full value at low temperature [see
  Fig.~\ref{fig:euo-raw}(a,b)], but recovers on warming towards
$T_{\rm C}$, so a fraction of muons may implant in some additional
state which depolarizes the muon very rapidly.

We note that a recent experiment \cite{sms} on SmS has shown evidence
for the formation of a bound magnetic polaron consisting of an
electron around the implanted muon, in which the electron localization
is stabilised by exchange energy.  This occurs in the paramagnetic
state in which a ferromagnetic droplet is localized in the
paramagnetic host.  A similar effect has been noted in EuS
\cite{storchak2} although the larger magnetization ensures it occurs
at temperatures $\gg T_{\rm C}$ \cite{storchak1} and the same will be
true in EuO in which the magnetization is even larger.  Therefore such
a muon-related polaron is not relevant for EuO in the studied
temperature regime.

\begin{figure}
\includegraphics[width=8.0cm]{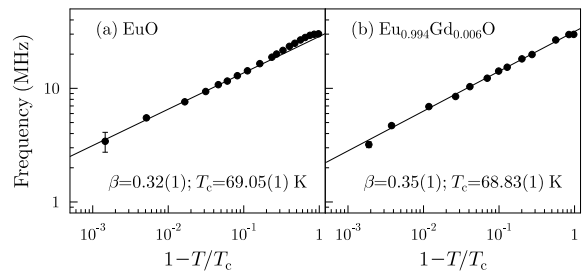}
\caption{Precession frequency extracted from $\mu$SR
  data as a function of temperature close to $T_{\rm C}$ for (a) EuO
  and (b) Eu$_{0.994}$Gd$_{0.006}$O.
\label{fig:freq}}
\end{figure}

The temperature dependence of the precession frequency for both EuO
and Eu$_{0.994}$Gd$_{0.006}$O was followed near $T_{\rm C}$ and the
results are plotted in Fig.~\ref{fig:freq}.  The fitted values of
$T_{\rm C}$ and the critical exponent $\beta$ are similar in each
case, though the value of $\beta$ is quite sensitive to the precise
value taken for $T_{\rm c}$.  Due to the difficulty in stabilising the
temperature better than $\approx 10$\,mK, we do not believe that the
difference between the two values of $\beta$ is significant.  They are
both close to 0.36--0.37 obtained using neutron scattering \cite{als}
and 0.38 obtained from a second order $\epsilon$ expansion for the
Heisenberg ferromagnet with dipolar interactions \cite{bruce}.

\begin{figure}
\includegraphics[width=8.0cm]{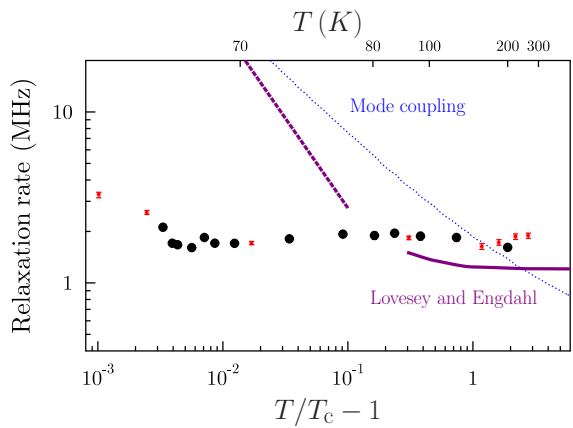}
\caption{(Color online.) Relaxation rate as a function of temperature
  in the paramagnetic regime for pure EuO (solid circles) and (b)
  Eu$_{0.994}$Gd$_{0.006}$O (dots).  The predicted relaxation rates
  for $T>T_{\rm C}$ according to Ref.~\onlinecite{lovesey3} (numerical
  calculation: thick solid line, purple; critical regime using
  experimental values of correlation length: thick dashed line,
  purple) and Ref.~\onlinecite{yaouanc} (dotted line, blue) are also
  shown.
\label{fig:relax}}
\end{figure}

Above $T_{\rm C}$, we observe simple exponential relaxation 
[Fig.~\ref{fig:euo-raw}(a,b)] with a
relaxation rate $\lambda$.
For zero-field relaxation of muons initially polarized
parallel to $z$, $\lambda$ can be written in terms of field-field
correlation functions using 
%\begin{equation}
$\lambda = \frac{\gamma_\mu^2}{2}
\int_{-\infty}^\infty {\rm d}t\, \left(\langle B_x(0)B_x(t) \rangle +
\langle B_y(0)B_y(t) \rangle\right)$ \cite{dalmas}.  
%\end{equation}
When each Eu spin component fluctuates, it produces a field
fluctuation via the resulting modulation of the dipolar and hyperfine
couplings.  Our measurements of the zero-field relaxation rate for the
EuO sample are plotted in Fig.~\ref{fig:relax} (a less complete set of
data for the Gd-doped sample is also shown).  There is a small rise in
$\lambda$ as the temperature is lowered towards $T_{\rm C}$ but apart
from this $\lambda$ remains just below $\approx 2$\,MHz for both
samples across the entire range studied.

These results can be compared with calculations on a localized
Heisenberg ferromagnet which have been performed with EuO in mind
\cite{lovesey3,yaouanc} (Fig.~4).  The theory of Lovesey and Engdahl
\cite{lovesey3} includes only the dipolar coupling and has been
evaluated for temperatures above $1.3T_{\rm C}$, assuming a muon site
of $\frac{1}{4}\frac{1}{4}\frac{1}{4}$.  Though underestimating the
observed experimental values, this theory does remarkably well in
providing a good estimate of the rough size of $\lambda$, the
discrepancy perhaps being due to neglecting the hyperfine
contribution.  It is known that critical fluctuations enhance the role
of the hyperfine coupling over the dipole coupling \cite{lovesey1}
because $B_{\rm dip}=0$ at the muon site in the ordered state and the
peak in the susceptibility is at ${\bf k}=0$.  Nevertheless, when the
dipolar calculation is extended into the critical regime (just above,
and very close to, $T_{\rm C}$) it predicts a divergence in $\lambda$
which is not observed.  An earlier mode-coupling approach
\cite{yaouanc} also predicts a very sharp increase in $\lambda$ on
cooling to $T_{\rm C}$ from about 0.3\,K above it; this is also not
observed in our data.  Magnetic polaron formation has been detected
using Raman scattering \cite{snow} in a narrow range ($\approx 20$\,K)
above $T_{\rm C}$. It may be that the formation of magnetic polarons
modifies the relaxation in this regime from that which would be
expected from theory, perhaps by providing an additional relaxation
channel for the muon which masks the critical slowing down predicted
by the theory and hence the absence of the divergence in $\lambda$.
We note that a similar absence of a divergence in $\lambda$ is
observed in EuB$_6$ \cite{brooks} in which magnetic polarons have been
found \cite{snow}.

In conclusion, we have identified the muon site in EuO and estimated
the hyperfine field.  Our results confirm long-range order which is
relatively insensitive to low doping of Gd.  The measured $\lambda$ in
the paramagnetic state agrees quite well with the theory of Lovesey
and Engdahl, but the available theories fail in the critical regime,
possibly due to magnetic polaron formation.

We thank EPSRC (UK) for financial support and
Alex Amato for experimental assistance.
Part of this work was performed at the Swiss Muon Source, Paul Scherrer
Institute, Villigen, Switzerland.


\end{document}